\documentclass[twocolumn]{aa}
\usepackage{graphicx}
\begin{document}
\bibliographystyle{bibtex/aa} 
\title{Assessing the galaxy population out to z$\sim$2}
\subtitle{using the Hubble Deep Field South}
\author{Theresa Wiegert\inst{1*}, 
         Du{\'\i}lia F. de Mello\inst{1,2,3},
         Cathy Horellou\inst{1}
          }

   \offprints{T. Wiegert\\
              \email{wiegert@physics.umanitoba.ca}\\
$*$Present address: Dept. of Physics and Astronomy,
University of Manitoba, Winnipeg, MB, R3T 2N2, Canada}

   \institute{Onsala Space Observatory, 
        Chalmers University of Technology, 
        SE-439 92 Onsala, Sweden
    \and
Laboratory for Astronomy and Solar Physics, 
Goddard Space Flight Center, Greenbelt MD20771, USA
\and
Catholic University of America Washington, DC 20064
             }

   \date{Received ; accepted}

   \abstract{In this work we use the Hubble Deep Field South (HDF-S) version 2 images
   to assess the galaxy population out to  $z \sim 2$. 
   We have used two methods of templates fitting of 
   the spectral energy distributions to obtain photometric 
   redshifts and classify the objects. The Bayesian photometric redshifts 
   gave better results when compared with 54 spectroscopic redshifts 
   available in the literature. 
   Analysis of the rest-frame colour distribution shows a bimodality out to  $
   z \sim 1.4$. 
   We separated our sample in a blue and a red population at $B-V=0.29$. 
   At low redshifts (0.2$<$z$<$0.6), $\sim$60\% of the galaxies have 
   $B-V>0.29$ whereas at higher redshifts $\sim$60\% of the galaxies are bluer
   than $B-V<0.29$. Although in low numbers, a population of early-type
   galaxies (or heavily obscured low redshift galaxies) is seen out to $z\sim$2.

   }

   \maketitle
\section{Introduction}

In spite of the observational and theoretical efforts which have
been made since Hubble (1926) classified galaxies according to their morphology, 
the epoch in which the Hubble types assembled and the process of formation are 
still unknown. Linking the high-$z$ population with the local universe is one of
the most difficult tasks in astronomy. The theoretical predictions such as
mass and size evolution are not easily accessible remaining a challenge for the
current instruments. Recently, Ravindranath et al. (2004) using the HST/ACS GOODS 
(Great Observatories Origins Deep Survey done with the Advanced Camera for Surveys with
the Hubble Space Telescope) images found no evidence of size evolution of disk galaxies 
over the redshift range 0.25$<z<$1.25. 
They find that the number densities of disk galaxies remains fairly constant over 
this redshift range. Bell et al. (2004) using the COMBO-17 survey (Classifying Objects by Medium-Band
Observations in 17 filters) found an increase in stellar mass on the 
red galaxies (i.e. early-types) by a factor of two since z$\sim$1. 
These results set strong constraints in the galaxy formation
theories such as the hierarchical scenario and the monolithic collapse. In the hierarchical 
scenario stars are formed continuously over a wide range of redshifts and galaxies are 
assembled via many generations of mergers of smaller subunits 
(e.g. White \& Frenk 1991,  Kauffmann et al. 1993 ). The monolithic collapse 
theory, on the other hand, predicts rapid star formation at very high redshift ($z >$ 2) followed 
by a steep decline in the star formation rates (Eggen et al. 1962, Jimenez et al. 1998).
If Ravindranath et al. (2004) results are confirmed, disks were already assembled
at z$\sim$1.25 and mergers occuring after that epoch would not affect their sizes.
Whereas if Bell et al. (2004) results are confirmed, early-type galaxies 
would have half of their today's masses at that epoch. Other independent analyses are needed in order to confirm these results.

In this work we use the Hubble Deep Field South (HDF-S) images which has the 
depth and resolution required to evaluate the galaxy population out to $z\sim$2. 
The population at higher redshifts has been addressed in the recent studies by 
Trujillo et al. (2004) and Labb\'e et al. (2003) 
using near-infrared observations of the HDF-S where they find a low number of massive
galaxies at z$\sim$3, i.e. 1.2 Gyr before the cosmic epoch which we are 
analyzing in this paper. 

In Section~2 we describe the data, photometric redshifts and $K$-corrections, 
in Section~3 we present the analysis of the rest-frame colour distribution. 
Conclusions are presented in Section~4.

\section{The data}

The HDF-S is the second deep field taken with HST (Williams et al. 2000)
reaching limiting magnitudes at the 10$\sigma$-level of 26.98, 27.78, 28.34, 27.72 
for F300W, F450W, F606W and F814W ($U$, $B$, $V$ and $I$ bands), 
respectively (Casertano et al. 2000). The WFPC2 field is 163$''\times183''$ or 
4.38 arcmin$^2$, covered by $4096\times4600$ pixels.

We have obtained the public version~2 of the WFPC2 HDF-S mosaicked images from 
the Space Telescope Science Institute's archive. The version~2 of the HDF-S is an 
improvement of the first release since the sky background was flattened, a new 
correction for scattered light in the U-band and more images were added. 
The noise level decreased and the sky is more homogeneous than in the first version.

The calibration of the raw data is described in Casertano et al. (2000). 
The four cameras of the WFPC2 were mosaicked into one image, after having been improved by 
dithering and drizzling. They were also resampled to the same pixel scale 
(0.0398 arcsec/pixel). The drizzling procedure combines various pointings, 
preserves photometry and resolution, and removes effects of geometrical distortion. 

The image depth fades towards the edges due to the variety of pointings, i.e.
the coverage in the edges is lower than in the middle of the separate parts of the
image. This means that the outer regions are less reliable, as well as the
regions around the cross that separates the four parts of the image.  The same was 
reported by Volonteri et al. (2000a) using the shallower version 1 data.

\subsection{Catalogues: Parameters and detections}\label{params}

We have catalogued the objects in the HDF-S using Source Extractor by 
Bertin \& Arnouts (1996; hereafter SE).
SE is especially useful for large images with a considerable amount of faint objects.
The output from SE is a catalogue of detections including photometry, as chosen by a 
set of input parameters.
The configuration file for SE has many input parameters which were tuned by visual
inspection to fit this set of data. Table~\ref{tab:se} lists some of the parameters chosen.

\begin{table*}
\centering
\begin{tabular}{ll|ll}
\hline
I-band Detection& & V-band Detection&\\
\hline
detect\_minarea & 18 & satur\_level&16\\
detect\_thresh &1.4 & mag\_zeropoint & 23.02\\
deblend\_mincount & 0.009 & gain & 680400\\
clean\_param& 0.7 & pixel\_scale & 0.0398 \\
\hline
\end{tabular}
\caption{Source Extractor parameters used. The right half shows values for
  detection in the $I$-band and the left half values for the photometry in the $V$-band.}
  \label{tab:se}
\end{table*}

We were able to detect objects in the limiting magnitudes of the HDF-S, however there were 
still some disagreement between over-merging and dividing of spirals. 
Since no single set of parameters gives a
completely satisfactory result, another detection method was used with a
larger deblending value. The objects which needed to be altered were visually
chosen and manually exchanged. 

The catalogue SE produced consists of 1310 detections. After the cleaning 
process and removing starlike objects, the final catalogue consists of 1142 objects. 
The magnitudes 
obtained are AB magnitudes, determined by the zero-point taken from 
Casertano et al. (2000). 
We have also used a magnitude-limited sample ($I<26.5$)
in the analysis, removing weak detections which yield bad photometric
redshifts. The magnitude-limited sample has 591 objects.

In order to avoid different numbers of detections in different filters,
we used the dual mode in SE. With the dual mode the same input parameters,
such as apertures and positions, were used and the catalogues were matched afterwards. 
In this case, we decided to use the high S/N $I$-band for detections. Four
runs of SE were executed - first a single mode run where
both detection and photometry were performed on the $I$-band. Then the
parameters for the photometry were changed for each of the other three bands,
to obtain the catalogues for the $U$-, $B$- and $V$-bands, respectively. 

Another advantage of using the dual mode feature is that the photometry is 
measured in the same aperture in every band, which is necessary for the
photometric redshift determination.

\subsection{Photometric redshifts, K-corrections}
Photometric redshifts were calculated using methods 
that fit spectral energy distributions (SED) with 
templates. 
We examined two public codes, {\tt Hyperz} by 
Bolzonella et al. (2000) 
and Bayesian Photometric Redshifts ({\tt BPZ}) 
by Benitez (2000).
A template fitting method compares templates at different redshifts with the four 
observed datapoints for each object. 
The best matching spectrum gives both the photometric redshift as well 
as the galaxy's spectral type. 

The difference between {\tt BPZ} and standard Maximum Likelihood methods like
{\tt Hyperz} lies in the addition of Bayesian probabilities, 
and the use of a prior for calculating them.
The prior is the central part of the {\tt BPZ} method. 
The prior used for this work is the one available in the public code, derived from the HDF-N, and the
$I$-magnitude is used as input for the prior determination (base magnitude M\_0).

We used four spectral type templates (Coleman, Wu \& Weedman 1980) 
for early-type, Sbc, Scd and irregular galaxies, as well as two blue starburst galaxy templates from 
Kinney et al. (1996) 
(SB3 and SB2, the first with a mean colour excess $E(B-V)$ between
0.25 and 0.35, the latter with $E(B-V)$ between 0.11 and 0.21).
The spectra are extended to the ultraviolet using a linear extrapolation and a cutoff at
912\AA, and to the near infrared using GISSEL synthetic templates. They
are corrected for intergalactic absorption following 
Madau (1995).

Deciding which photometric redshift approach is the best one is not a simple
task since the number of spectroscopically confirmed galaxies 
in the HDF-S only reaches 54 so far
(Cristiani et al.  2000, Franceschini et al. 2003, Sawicki \& Mallen-Ornellas 2003). 
In both softwares, the best result is given when using the
low-redshift empirical library of SEDs instead of a synthetical template set
that takes galaxy evolution into account.

In some cases the spectrum of a galaxy has no good
equivalent in the template library, but will still be assigned a redshift
corresponding to the closest match, even though it might be far from the
observed colours and thus the real redshift. 
Why then not extend the template library to include more templates, 
to cover all possible galaxy types?  
The reason is that the method considers
all the templates equal in status, which means that a larger template set
increases the number of colour/redshift degeneracies 
(Benitez, 2000).

\begin{figure}
\includegraphics[width=8cm]{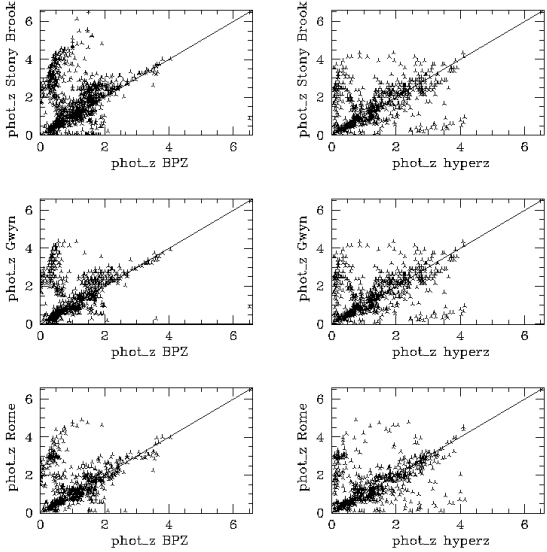}
\caption{Comparison between our photometric redshifts obtained with {\tt Hyperz} 
  and {\tt BPZ} with other catalogues: 
the Stony Brook group, Gwyn and the Rome group  Volonteri (2000b).
}
\label{fig1}
\end{figure}

In Fig.~\ref{fig1} we compare our 
{\tt Hyperz} and {\tt BPZ} photometric redshifts with 
three others that are available in the literature: 
(Stony Brook\footnote{\tt{www.ess.sunysb.edu/astro/hdfs/wfpc2/wfpc2.html}},
Gwyn\footnote{\tt{astrowww.phys.uvic.ca/grads/gwyn/pz/hdfs/}} 
and Rome (Volonteri et al., 2000b).
Since the catalogues were generated with different SE input parameters, 
we matched the catalogues and
considered only the objects in common. 
Overall our {\tt BPZ}  photometric redshifts agree better 
with the ones obtained by other groups 
and thus became our choice for further analysis. 
The major disagreement is in the upper-left part of the plots,
where our values of photometric redshifts are much lower than
the ones by the other authors.  
We have checked those objects and found that they are mostly very faint objects 
close to the detection limit which will be filtered out in a magnitude-limited sample. 
Moreover, 
Benitez (2000) 
has tested {\tt BPZ} using the HDF-N which has $\sim 200$ spectroscopic redshifts available 
with good results, which encouraged us to use his method.
Recently, 
Mobasher et al. (2004a)
have also used this method for the GOODS survey which has a larger number of 
spectroscopically confirmed galaxies in the HDF-N and Chandra Deep Field South.

\begin{figure}
\includegraphics[width=8cm]{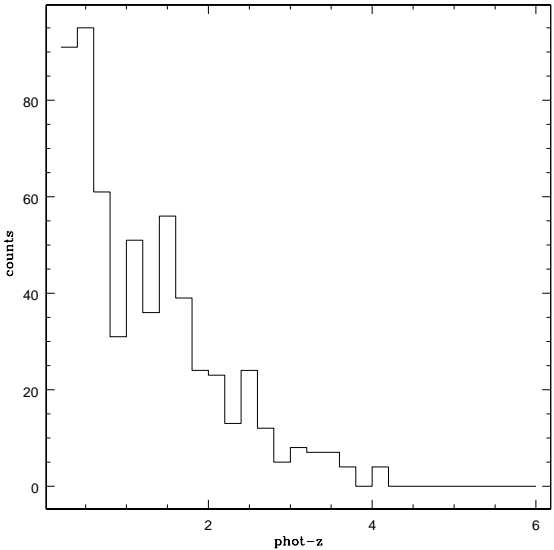}
\caption{Distribution of {\tt BPZ} photometric redshifts of 
the magnitude-limited catalogue ($I<26.5$).}
\label{fig2}
\end{figure}

\begin{figure}
\includegraphics[width=8cm]{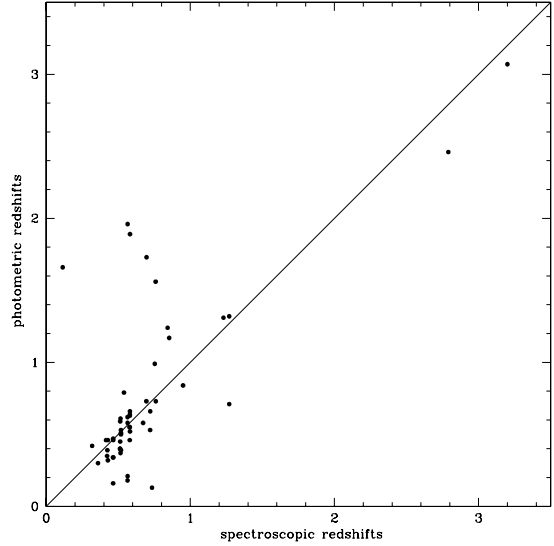}
\caption{Comparison between photometric redshifts and spectroscopic redshifts.}
\label{fig3}
\end{figure}

Fig.~\ref{fig2} shows the photometric redshift distribution 
of the magnitude-limited catalogue (I$<$26.5) where the majority of objects
are found at z$<$2. Unfortunately, the number of spectroscopically confirmed
redshifts in the HDF-S is not as high as in the HDF-N. We found 54 redshifts
within the WFPC2 area we are using. 
Sawicki \& Mallen-Ornellas (2003)
have 55 spectroscopic redshifts, but six of these are
double detections (i.e. the resolution of the ground-based observations was  
not good enough, making two galaxies share the same
redshift) and three are not within the boundaries of the WFPC2 area we are using. 
Fig.~\ref{fig3} shows the comparison between our {\tt BPZ} 
photometric redshifts and the spectroscopic redshifts.
We concluded that our photometric redshifts are overall in good agreeement 
with the spectroscopic redshifts. 
However, there are very few objects with spectroscopic redshifts $z>1.5$ 
available in the literature. We will limit our analysis to the 
range 0.2$<z<$1.4 where we have enough spectroscopic redshifts to test 
the photometric redshifts. 
There are seven objects with photometric redshifts significantly
different from their spectroscopic redshifts (Table~\ref{tab:discordant}). 
We note that all but one of the discordant redshift objects were assigned an
early-type template. We believe that wrong redshifts were assigned due 
to extinction corrections which were not applied in our spectral classification. 
Taking into consideration that we only had 4 colours, the template
fitting method to obtain redshifts worked remarkably well. This is an important result 
considering how time consuming deep $HST$ observations are. 

\begin{table}[!htp]
\centering
\begin{tabular}{rllcll}
\hline
Id  &   $Phot-z$  &  $Spect-z$  &  Type & I$_{AB}$  & $B-V$\\
\hline
 306 &  0.13 &   0.7342  &  1   &   20.0  &  0.7041\\
 594 &  1.96 &   0.5648  &  2   &   23.0  &  0.4517\\

 638 &  1.89 &   0.5817  &  2   &   23.0  &  0.4794\\
1091 &  1.56 &   0.7594  &  2   &   23.8  &  0.4871\\
1264 &  1.73 &   0.6963  &  2   &   25.0  &  0.4805\\
 632 &  0.71 &   1.27    &  2   &   22.9  &  0.4319\\
1101 &  1.66 &   0.1148  &  4   &   23.8  &  0.2460\\  
\hline
\end{tabular}
\caption{Catalogue number, {\tt BPZ} photometric redshift, 
spectroscopic redshift, spectral type (1=E, S0, Sa, 2=Sbc), 
I$_{AB}$ magnitude, $B-V$ colour.}
\label{tab:discordant}
\end{table}

The rest-frame magnitudes were calculated using 
$K$-corrections and distance moduli provided by T. Dahl\'en (2003, priv. comm.), 
covering redshifts up to $z=7$ for the six spectral types and a cosmology with 
$H_0=70$ km~s$^{-1}$Mpc$^{-1}$, 
$\Omega_{M}=0.3$ 
and $\Omega_{\Lambda}=0.7$.

\subsection{Spectral types}

Since the first Hubble Deep Field images (Williams et al. 1996) several authors have attempted 
to classify the morphology of faint galaxies (e.g. Abraham 
et al. 1996, van den Bergh et al. 1996, Corbin et al. 2001, Menanteau et al. 2001, van den 
Bergh et al. 2002). Automatized methods of classification, such as the one 
which measures structural parameters (e.g. Conselice 2003), produce valuable results 
(e.g. Conselice et al. 2004, Mobasher et al. 2004b). 
However, instead of using the concentration of stellar light and its asymmetric distribution 
to classify the galaxies in the HDF-S we chose to use the spectral types which were 
obtained from the template fitting in the photometric redshift technique as a morphology 
indicator. This latter method keeps the classification directly attached to the 
photometric redshift estimates. Moreover, the spectral types are associated with the
spectral energy distribution which reflects the physical properties of the objects, 
such as star formation history. 
We have randomly selected a sub-sample (15\% of the total sample) for visual 
inspection. A gallery of typical galaxies is available online\footnote 
{\tt{www.oso.chalmers.se/$\sim$theresaw/Deep/gallery.html}}. 
The gallery is divided
in spectral types and redshift bins. Galaxies with 0.2$<z<$1 which were assigned an E-Sa 
spectral template indeed resemble classical early-type galaxies whereas more distant galaxies 
(1$<z<$2) are red objects too faint to be visually classified. Galaxies which were 
assigned Sbc and Scd spectral templates show the presence of a disk in the low redshift bins
and star formation. The ones assigned irregular and starburst (SB2 and SB3) templates 
have clearly peculiar morphologies in all redshifts. Some SB3 galaxies with 0.2$<z<$0.5 are 
less peculiar showing disks with strong knots of star formation.  
We have not attempted to correct for extinction and we are aware of the effects that this 
can cause. For instance, the red colours in the early type galaxies at high$-z$ might be
due to dust and not due to an old population. However, considering the good agreement that we
have obtained in the photometric redshifts, only a few objects would require
strong extinction correction which would change the spectral type assigned.

\begin{figure}[!htp]
\includegraphics[width=8cm]{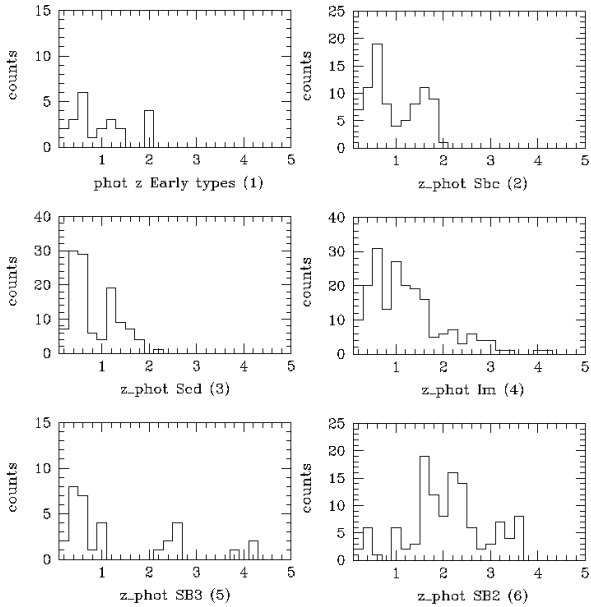}
\caption{Distribution of {\tt BPZ} photometric redshifts for each of the galaxy types.}
\label{fig:fig4}
\end{figure}

The redshift distribution for each spectral type 
(Fig.~\ref{fig:fig4}) clearly
shows that 
the number of early-type galaxies (E, S0, Sba) decreases with redshift. 
The amount of irregular and starburst galaxies, on the other hand, increases with redshift. 
There is a high amount of irregulars all through the redshift range covered. 
However, at high redshifts, only irregulars and starbursts 
are present. The irregulars peak at $0.75 < z < 1.0$, and starbursts (SB2) at $z>1.0$. 
This is in agreement with
Volonteri et al. (2000b) 
who found that early-type galaxies show a steep decrease in number counts in the
faintest bin. 

\begin{table*}[!htp]
\centering
\begin{tabular}{c|llll|llll}
\hline
 &&& Red Group &&&& Blue Group & \\
 catalogues& mean & $\sigma$ & median & objects & mean & $\sigma$ & median & objects \\
\hline
 $a$ & 0.458 & 0.247 & 0.400 & 440 & 0.125 & 0.119 & 0.161 & 700\\
 $b$ & 0.429 & 0.138 & 0.385 & 317 & 0.166 & 0.095 & 0.186 & 274\\
 $c$ & 0.440 & 0.131 & 0.404 & 257 & 0.153 & 0.087 & 0.170 & 334\\
 $d$ & 0.424 & 0.128 & 0.379 & 174 & 0.194 & 0.057 & 0.197 & 123\\
\hline
 $d2$ & 0.399 & 0.065 & 0.405 & 107 &&&&\\
\hline
\end{tabular}
\caption{Mean, $\sigma$ and median values of the $B-V$ colours listed for the full catalogue and its subsamples. 
$a$ is the full catalogue,
$b$ is the magnitude-limited catalogue, 
$c$ is the full catalogue limited in redshift ($0.2 < z < 1.2$) 
and $d$ is the magnitude-limited
catalogue limited in redshift. 
$d_2$ is the red group from $d$ without the red objects.}\label{tab:mean}
\end{table*}

\section{Discussion}

\begin{figure}
\includegraphics[width=8cm]{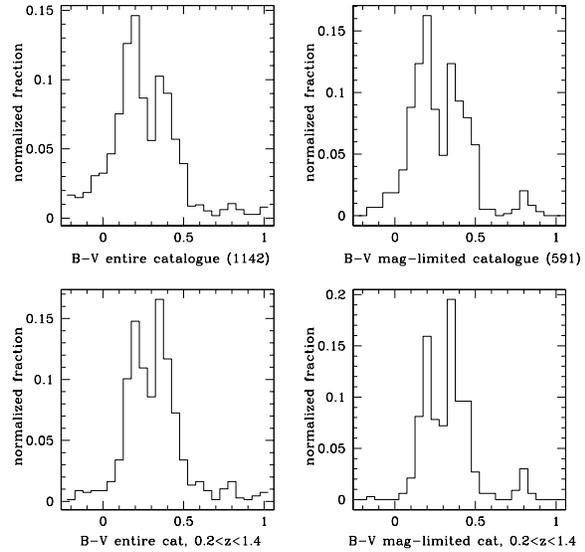}
\caption{Colour distribution for all galaxies in the sample and for the magnitude-limited 
sample.}
\label{fig5}
\end{figure}

The colour distribution for all galaxies in the sample and for the magnitude-limited sample is 
shown in Fig.~\ref{fig5}. For comparison, local late-types (Scd) and irregular galaxies have $B-V$=0.15--0.48
($B-V$=0.27--0.60 Johnson filters), 
whereas earlier types have $B-V=$0.73--0.80 (B-V=0.85--0.92) (Liu \& Green 1998). 
Therefore, we are clearly detecting a larger number
of star-forming galaxies in all redshift bins. 
Two distinct populations of objects are clearly seen in Fig.~\ref{fig5}, one
which is redder 
than $B-V=0.29$, and one which is bluer than that.
The bimodality in the colour distribution of galaxies has recently been seen 
by Strateva et al. (2001) using the Sloan Digital Sky Survey, Hogg et al. (2002) 
for the local universe ($z<0.22$) and Bell et al. (2004) out to $z \sim 1$. 
For our sample, the bimodality is actually still visible out to a redshift of $z \sim 1.4$.
Similar analysis using the ESO Imaging Survey (EIS)
has shown a bimodality 
for redshifts out to $z \sim 1.6$ for a much larger area in the sky (J. Andersson
private communication).
The colours of galaxies reflect their star formation
history, therefore a bimodality in colours suggests that they have gone
through different evolutionary paths.

\begin{figure}[!htp]
\includegraphics[width=8cm]{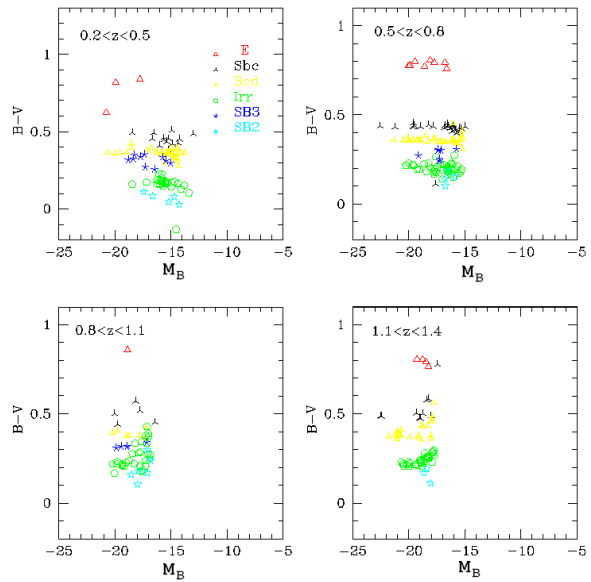}
\caption{Rest-frame colour-magnitude diagram for four redshift bins.}
\label{fig:fig6}
\end{figure}

We chose to do the analysis on the magnitude-limited sample ($I<26.5$) instead of the entire catalogue which has limiting I-magnitude 28.6. 
The disadvantage in using the former sample is the smaller number of objects. 
However, the latter sample would include spurious faint outliers which were still present in the redshift-limited catalogue.
A consequence of this limitation is that objects are being dropped from the sample. 
The colour-magnitude diagrams are shown in Fig.~\ref{fig:fig6}.
Comparing Fig.~\ref{fig:fig6} with  Fig.~\ref{fig:fig4} we notice that 
there are many more starbursts and irregulars in the entire sample than 
in the magnitude-limited sample.

In Table~\ref{tab:mean} we give the average mean, $\sigma$ and median of the $B-V$ colours for
the blue and red sample using the full catalogue and the magnitude-limited catalogue. 
We also include the same for the redshift-limited catalogue. The average colour of the
blue catalogue is $B-V=0.125 \pm 0.119$, whereas the mean value for the magnitude and redshift-limited
sample is $B-V=0.194 \pm 0.057$ showing a low dispersion in colours. The average colour of the 
red catalogue is strongly affected by a small group of redder objects with $B-V>0.7$ present in the top of the plots, 
which is still noticeable in the highest redshift bin of Fig.~\ref{fig:fig6}. 
The red group has a larger standard deviation ($\sigma$) than the blue, caused by these redder objects. If we remove them from the red sample (d2 in 
Table~\ref{tab:mean}), the dispersion in the colour is comparable with the one for the blue group. 
The red group is on average a factor of 2 redder than the blue 
group (sample d in Table~\ref{tab:mean}). For comparison, the u$^{*}$--r$^{*}$ colours of the red group in 
Strateva et al. (2001) is also $\sim$ 2 times redder than the blue group (from Fig. 2 in their paper) while 
in Bell et al. (2004) the $U-V$ colours of the red population can be as high as three times the colours of the blue 
population (From Fig. 1 in their paper). We find a factor of $\sim 3$ (or higher) if we 
use the reddest objects instead of average values. However, there are very few of these red objects and a larger field is needed to improve the statistiscs.

The median redshift for the blue population is $z$=0.65 whereas for the red population 
is $z$=0.57. One interesting result is the fact that the redder objects have median $z$=0.61 and the 
bluer objects have $z$=$0.3-0.4$. These red objects probably have an older stellar
population and/or are heavily extinct, whereas the blue ones are galaxies experiencing
star formation. There are also a few very blue objects with $B-V<0.1$. These are 
strong starburst galaxies at low redshifts with $B-V$ bluer than local irregular galaxies. 


\begin{figure}[!htp]
\begin{center}
\includegraphics[width=8cm]{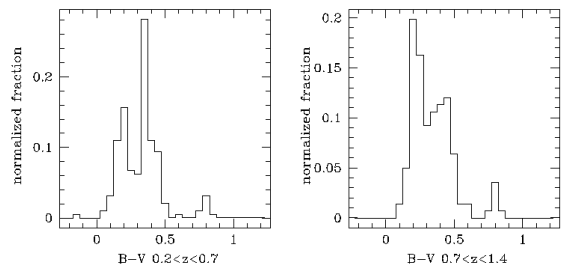}
\caption{Histograms of the magnitude-limited catalogue, divided into two redshift bins.}\label{fig:fig7}
\end{center}
\end{figure}

The bimodality of colours is seen at low and high redshifts (Fig.~\ref{fig:fig7}).
In the first bin, at lower redshifts, $\sim$60\% of the galaxies have 
($B-V>0.29$) whereas at higher redshifts $\sim$60\% of the galaxies are bluer
than $B-V<0.29$. The increase in blue galaxies at higher redshifts is 
clearly due to the irregulars and starbursts as shown in Fig.~\ref{fig:fig6}. 
We can also see in Fig.~\ref{fig:fig6} that the starburst galaxies of type 
SB2 are generally bluer than the irregular galaxies and are also contributing
to the blue colors. The starbursts SB3, on the other hand, are redder and
not all of them have $B-V<0.29$. 

Although in lower numbers ($\sim$5\%) than late-type galaxies, early-type 
galaxies are also seen out to z$\sim$2. 
This has important implications in galaxy evolution since it 
might imply 
that early-type galaxies are already in 
place at z$>$1. The question of when these galaxies are assembled will soon be answered when larger deep fields in multiwavelengths will be analysed. 
A caveat to be considered is the role of dust in galaxy evolution. Galaxies 
with large amount of dust such as submm galaxies (e.g. Eales et al. 2000) and 
$K$-luminous galaxies (Daddi et al. 2004) would not be easily detected 
in the images due to extinction and in case they were detected they would be too faint and lower than the magnitude limit adopted.
However, we cannot exclude the possibility that these redder galaxies are being heavily obscured by dust. Combining UV-optical with infrared is then essential (Conselice et al. 2003). Soon deep infrared images, 
such as the recent NICMOS Ultra Deep Field images, and future Spitzer images combined with HST UV-optical will be ideal for the study of galaxy evolution.

\section{Summary}

We have analysed the HDF-S $U,B,V,I$ images in order to assess the
galaxy population at $z<2$. 
We summarize our main results as follows:  

\begin{enumerate}

\item The template fitting method, BPZ, gives better
photometric redshifts than other redshift estimates and 
agrees well with the 54 spectroscopic redshifts available in
the literature. 
Some disagreement persisted for faint objects close to the
detection limit, and this conducted us to produce and perform the
analysis on a magnitude-limited catalogue of 591 objects with $I < 26.5$. 

\item The spectral types obtained from the template fitting were used as morphological indicator. 
The quality of the automated spectral type assignments was carefully checked by visual inspection. A gallery with typical types was produced.

\item A bimodal distribution of $B-V$ colours was found out to $z\sim 1.4$. 
At low redshifts (0.2$<$z$<$0.6), $\sim$60\% of the galaxies have 
($B-V>0.29$) whereas at higher redshifts $\sim$60\% of the galaxies are bluer
than $B-V<0.29$. The blue population has $<B-V>=0.194 \pm 0.057$ and the red population
has $<B-V>=0.399 \pm 0.065$. There is also a group of red objects with $B-V > 0.7$
present in all redshift bins whereas the bluest objects with $B-V < 0.2$ are
seen only in the lowest redshift bins ($0.2<z<0.5$).

\item Galaxies with spectral types typical or irregulars and starbursts are more numerous at higher redshifts ($z>0.8$) than
Sbc, Scd and early-types. 

\item The redder objects have median $z$=0.61 whereas the 
bluer objects have $z$=$0.3-0.4$.

\item Although in lower numbers (5\%), a population of early-type
galaxies (or heavily obscured low redshift galaxies) is seen out to $z\sim$2. 

\end{enumerate}

The results presented here suggest that galaxies of all spectral types 
(E-Sa, Sbc, Scd, Irr and starbursts) are seen out to $z\sim 1.4$ and
are separated in a blue and a red population. This bimodality in colours suggest
that galaxies have different star formation histories which could lead
to the different Hubble types seen in the local universe. Multiwavelength deep
observations of large areas are needed in order to establish the epoch of
galaxy formation.

\begin{acknowledgements}
Support for this work was provided by the Swedish National Space Board {\it Rymdstyrelsen} through grant 1629018. Part of the work was supported by the Swedish Research Council {\it Vetenskapsr\aa det}. We are grateful to Tomas Dahl\'en for his help on the $K$-corrections.
\end{acknowledgements}



 \end{document}